\documentclass[a4paper,11pt]{article}
\usepackage{pos}
\usepackage{booktabs}
\usepackage{bm}
\usepackage{graphicx}
\usepackage{braket}
\usepackage{makecell}
\newcommand{\rev}[1]{{#1}}

\renewcommand{\logo}{\relax}

\title{Thermal equilibrium curves of accretion disks driven by magnetorotational instability}
\ShortTitle{Thermal equilibrium curves of accretion disks}

\author*{Shigenobu Hirose}

\affiliation{Center for Mathematical Science and Advanced Technology, \\Japan Agency for Marine-Earth Science and Technology,\\
  Yokohama, Kanagawa 2360001, Japan}

\emailAdd{shirose@jamstec.go.jp}

\abstract{
Analogous to the HR diagram for stars, the thermal equilibrium curve encodes the thermodynamics of accretion disks by expressing the local balance between heating---primarily via viscous dissipation---and cooling---typically through radiative transfer. These curves are commonly plotted as surface density versus effective temperature. When an S-shaped locus appears, local annuli become bistable, and limit-cycle oscillations arise when the external mass-transfer rate falls within an unstable band. This behavior underpins the disk instability model for recurring outbursts in cataclysmic variables. 
This paper reviews first-principles thermal equilibrium curves for accretion disks driven by magnetorotational instability (MRI), with emphasis on dwarf novae. Unlike the parameterized $\alpha$-viscosity approach, the curves are obtained by solving the governing equations with radiation magnetohydrodynamics simulations, thereby reproducing S-shaped loci without prescribing $\alpha$. The disk instability in dwarf-nova systems and the physical origin of angular-momentum transport (shear stresses) are also briefly reviewed. Notes on the stability of radiation-dominated accretion flows are included in the Appendix.
}

\FullConference{87th Fujihara Seminar: The 50th Anniversary Workshop of the Disk Instability Model in Compact Binary Stars (DIM50TH2025)\\
22-26 September 2025\\
Tomakomai, Japan\\}

\begin{document}
\maketitle

\section{Introduction}
\begin{figure}[ht]
    \centering
    \includegraphics[width=0.75\linewidth]{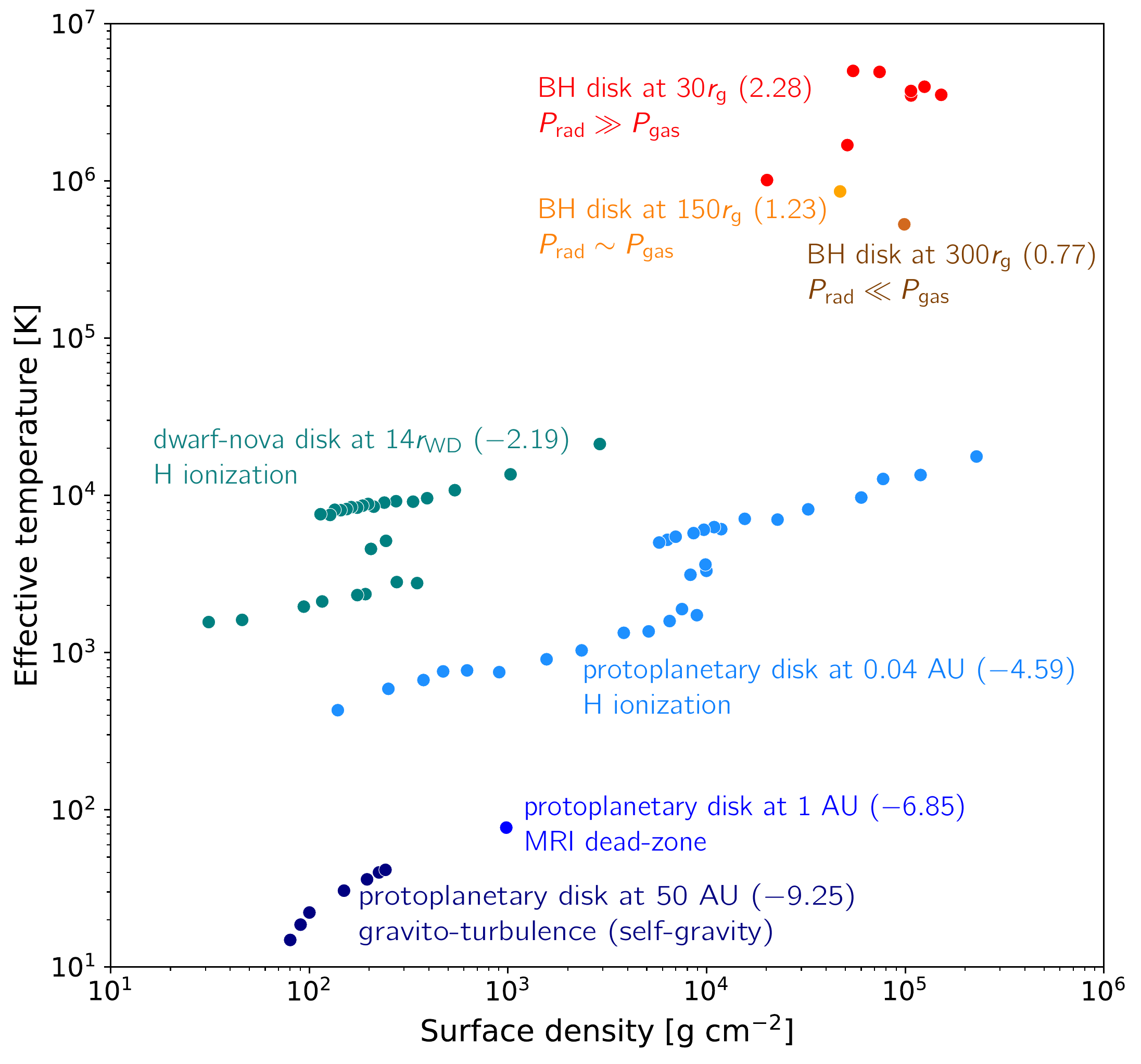}
    \caption{Thermal equilibrium curves in the $\Sigma$--$T_{\rm eff}$ plane for several disk classes (colors denote classes). Values in parentheses are $\log_{10}\Omega ~[\mathrm{s}^{-1}]$. \rev{BH disk at $30 r_\text{g}$ ($P_{\rm rad}\gg P_{\rm gas}$): Refs.~\cite{Hirose_2009a,Hirose_2009b,Blaes_2011}. BH disk at $150r_\text{g}$ ($P_{\rm rad}\sim P_{\rm gas}$): Refs.~\cite{Krolik_2007,Blaes_2007}. BH disk at $300r_\text{g}$ ($P_{\rm rad}\ll P_{\rm gas}$): Ref.~\cite{Hirose_2006}. dwarf-nova disk at $14r_{\rm WD}$: Ref.~\cite{Hirose_2014}. Protoplanetary disk at 0.04 AU (hydrogen ionization): Ref.~\cite{10.1093/mnras/stv203}. Protoplanetary disk at 1 AU (MRI dead zone): Ref.~\cite{Hirose_2011}. Protoplanetary disk at 50 AU (self-gravity): Refs.~\cite{10.1093/mnras/stx824,10.1093/mnras/stz163}. Detailed parameters used to calculate the gravitational radius $r_\text{g}$ and the white dwarf radius $r_\text{WD}$ are given in the respective references.} All data and scripts needed to reproduce this figure are available at Ref.~\cite{RMHD_TEs}.}
    \label{fig:TEcurve_all}
\end{figure}

Figure \ref{fig:TEcurve_all} is an overview of thermal equilibrium curves in the surface density--effective temperature ($\Sigma$--$T_{\rm eff}$) plane across several disk classes. These curves, compiled from the papers cited in the caption, cover black hole (BH) accretion disks, dwarf-nova disks, and protoplanetary disks. All cases use a common radiation magnetohydrodynamics (MHD) shearing-box framework; the control parameter is the angular velocity $\Omega$, spanning $10^{-9}$--$10^{2}\ \mathrm{s}^{-1}$. Because $\Omega$---and hence the local vertical gravity---varies by many orders of magnitude, the effective temperature ranges from $10$ to $10^{6}~\mathrm{K}$. 
Consequently, as the sampled $\Omega$ varies across disk classes, the dominant physics changes: self-gravity (outer protoplanetary disks), hydrogen ionization (dwarf-nova disks), and radiation pressure (inner BH disks).

The remainder of the paper focuses on the dwarf-nova disk case, which exhibits a characteristic S-shaped thermal equilibrium curve. Brief comments on the thermal stability of radiation-dominated BH disks are given in Appendix~\ref{sec:notes_prad}.

\section{Brief review of disk instability and thermal equilibrium curves}\label{sec:section1}

Relaxation oscillation provides the basic picture of dwarf-nova outbursts. This was first noted by Ref.~\cite{wesselink1939stellar-940}, who saw a similarity between dwarf-nova light curves and the discharge current in a neon tube oscillator. The neon tube has an S-shaped voltage--current characteristic with dark/glow bistability, and the system follows a limit cycle producing repeated flashes. It was also correctly pointed out in that work that hydrogen ionization is key to the relaxation process. However, this was long before the accretion--disk picture, so the mechanism was placed in the stellar envelope rather than in a disk.

\begin{figure}[ht]
    \centering
    \includegraphics[height=5cm]{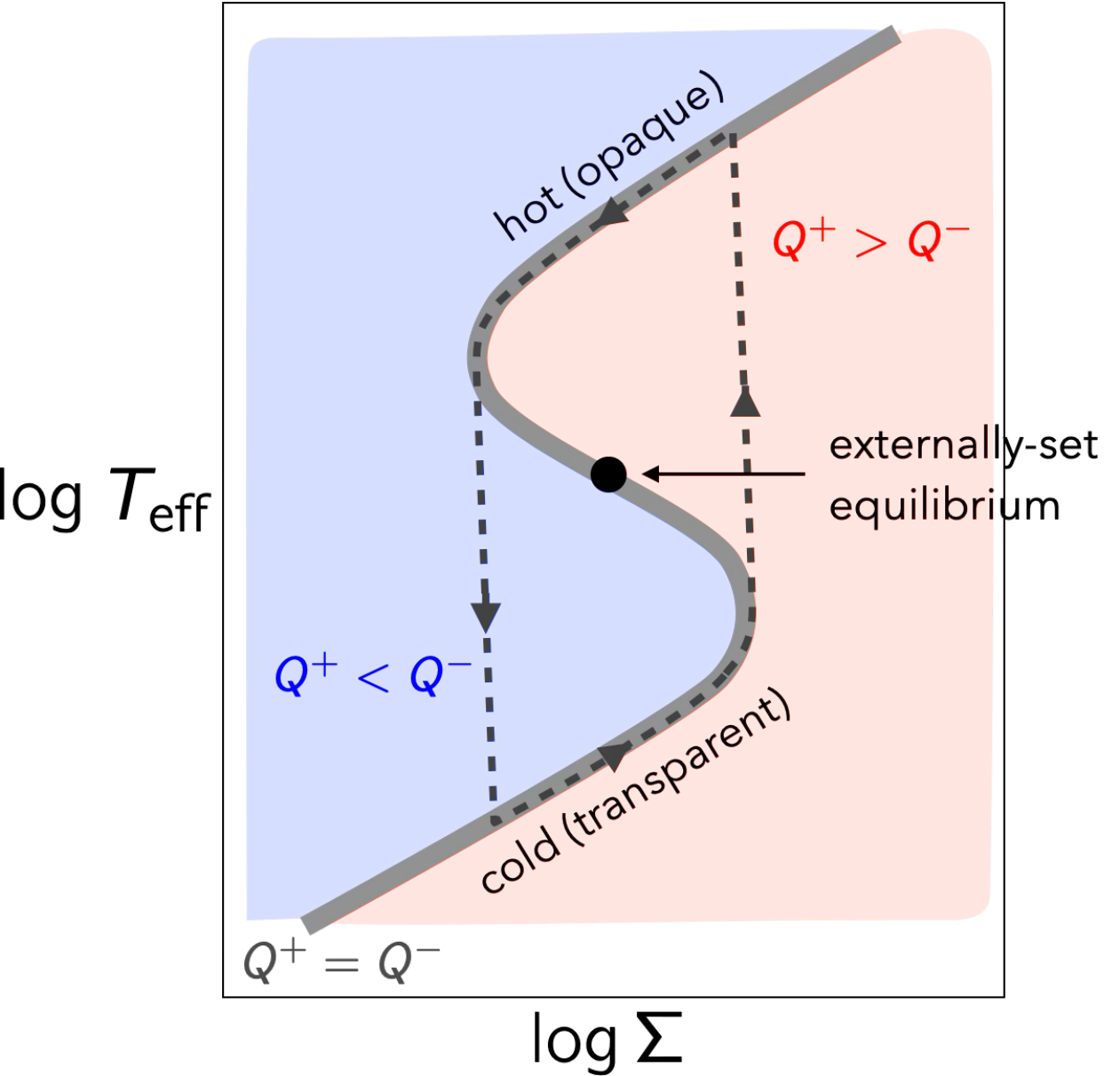}
    \includegraphics[height=5cm]{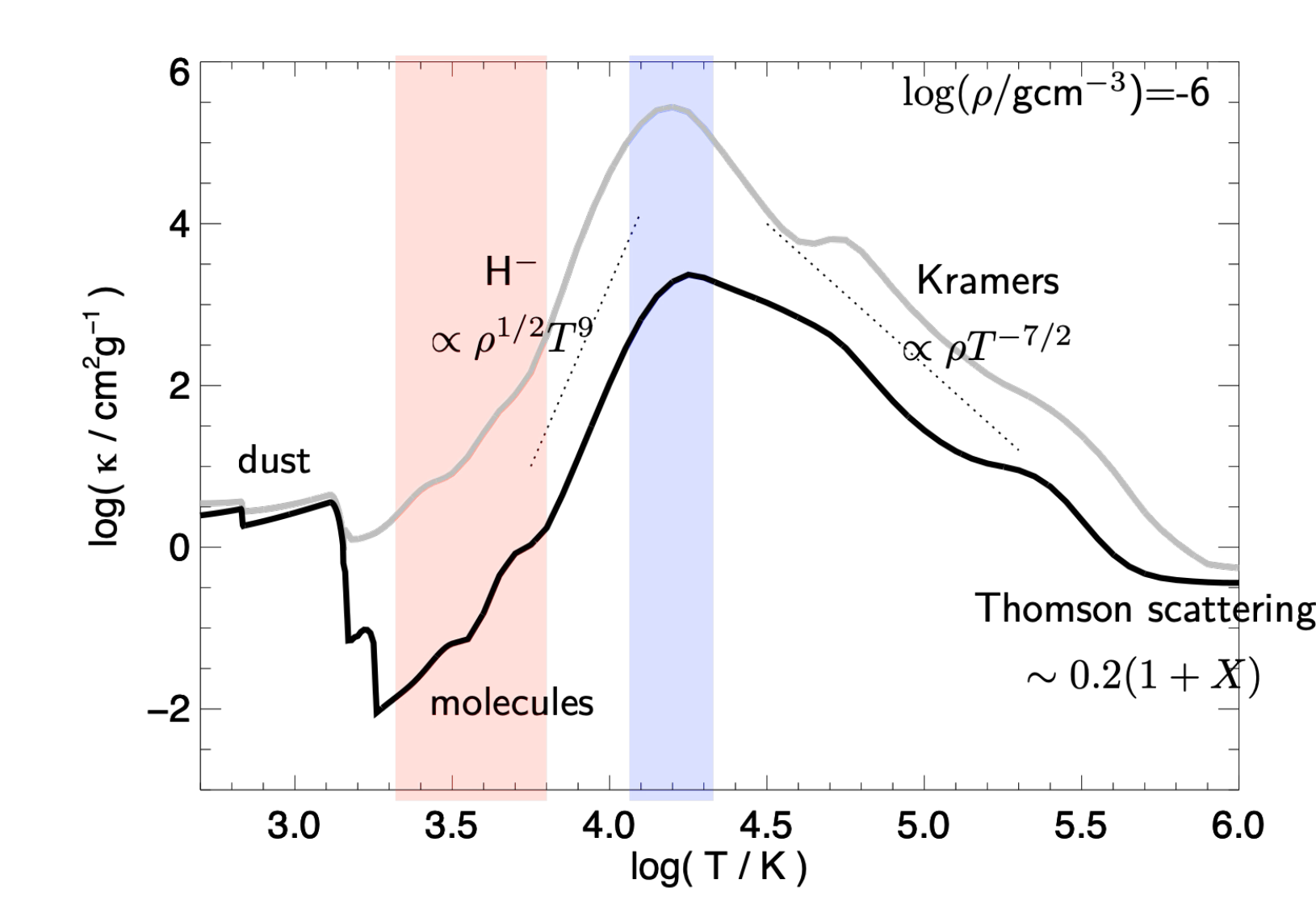}
    \caption{Left: Schematic S-shaped thermal equilibrium curve in the $\Sigma$--$T_{\rm eff}$ plane. Right: Rosseland-mean opacities (black) near hydrogen--ionization temperatures at a density $\rho=10^{-6}\ \mathrm{g\,cm^{-3}}$~\cite{optab14}.}
    \label{fig:S-curve}
\end{figure}

\rev{The idea that repeated outbursts of dwarf novae are driven by instabilities within the accretion disk---the disk instability model (DIM)---was first proposed in Ref.~\cite{1974PASJ26429O}.} The DIM was then developed by many authors in the 1980s; for historical context, see comprehensive reviews, e.g., Ref.~\cite{LASOTA2001449}. In this picture, the relaxation oscillation occurs on an S-shaped thermal equilibrium curve in the $\Sigma$--$T_{\rm eff}$ plane (Fig.~\ref{fig:S-curve}, left). When the operating point set by the external accretion rate falls on the unstable branch with negative slope, the system cannot remain there and instead jumps between cool and hot states, producing recurrent outbursts.

The thermal equilibrium curve $T_{\rm eff}=T_{\rm eff}(\Sigma; r)$ is local: at each radius $r$, it is set by balance between viscous heating $Q^+$ and radiative cooling $Q^-$,
\begin{align}
\rev{\underbrace{2\sigma_\mathrm{B} T_{\rm eff}^4}_{Q^-}
= \underbrace{\tfrac{3}{2}\,\Omega(r)\!\int\! \left(-w_{r\phi}\right)\,dz}_{Q^+}
= \tfrac{3}{2}\,\Omega(r)\,\alpha \!\int\! p\,dz
\approx \tfrac{3}{2}\,\Omega(r)\,\alpha\,\Sigma\,RT_{\rm mid},}
\end{align}
where $\Omega(r)$ is the angular velocity, $R$ is the gas constant per unit mass, and the $\alpha$ prescription is used \rev{($\int (-w_{r\phi})\,dz=\alpha\int p\,dz$)}. This yields a formal thermal equilibrium curve, 
\begin{align}
\rev{
\frac{\sigma_\mathrm{B}\,T_{\rm eff}^4}{RT_{\rm mid}(T_{\rm eff})}\;\sim\;\alpha\,\Omega(r)\,\Sigma,}\label{eq:S-curve}
\end{align}
if $T_{\rm mid}$ is regarded as a function of $T_{\rm eff}$.

Equation~\eqref{eq:S-curve} implies that a negative slope of $T_{\rm eff}(\Sigma)$ appears when $T_{\rm mid}(T_{\rm eff})$ grows faster than $T_{\rm eff}^4$, equivalently when $T_{\rm eff}$ depends only weakly on $T_{\rm mid}$. This typically occurs near hydrogen ionization where opacity varies sharply (Fig.~\ref{fig:S-curve}, right). Two effects make $T_{\rm eff}$ weakly dependent on $T_{\rm mid}$: (i) around $\sim 5{,}000~\mathrm{K}$ (red band), the strong $T$-dependence of the Rosseland-mean opacity $\kappa(T)$ weakens the $T_{\rm eff}$--$T_{\rm mid}$ sensitivity (under the optically thick diffusion scaling $\sigma_{\rm B}T_{\rm eff}^4 \sim T_{\rm mid}^4/[\kappa(T_{\rm mid})\Sigma]$) \rev{as first noted in Ref.~\cite{10.1143/PTP.61.1307}}; (ii) around $\sim 15{,}000~\mathrm{K}$ (blue band), an opacity bump can drive convection, which transports heat efficiently and flattens the $T_{\rm eff}$--$T_{\rm mid}$ relation. Together, these mechanisms yield two distinct negative-slope (thermally unstable) segments on the S-curve in both DIM (e.g., \rev{Fig.~2 in Ref.~\cite{1985PASJ371M}}) and radiation MHD simulations~\cite{Hirose_2014,10.1093/mnras/stv203} (see also Figs.~\ref{fig:Teff-Tmid} \rev{and \ref{fig:SC_alpha}}).


Modeling outburst decays as viscous diffusion on the outer-disk scale implies $\alpha_{\rm hot}\simeq 0.1$--$0.2$ on the hot branch. In contrast, reproducing outburst amplitudes and recurrence times suggests $\alpha_{\rm cool}\simeq 0.01$--$0.03$ on the cool branch (about one tenth to one quarter of the hot-branch value). These inferences, \rev{first pointed out in Ref.~\cite{Smak_1984},} are robust and have been confirmed by many authors (e.g., Ref.~\cite{Kotko_2012}). However, because $\alpha$ is a tunable parameter within the $\alpha$-prescription, such fits alone do not explain (i) what sets the absolute value of $\alpha$ and (ii) why its value differs between the hot and cool branches. To address these questions, the shear stresses from first principles are explored in the following sections.

\section{Origin of shear stresses --- magnetorotational instability (MRI)}

\subsection{Hydrodynamic Keplerian flow is stable}
Candidates for angular-momentum transport in Keplerian flows were discussed in Ref.~\cite{Shakura1973}, the original $\alpha$--disk paper. It was argued that magnetic fields could plausibly provide stresses, but the possibility of hydrodynamic turbulence was also mentioned, citing experiments that showed subcritical transition in Rayleigh--stable flows at high Reynolds numbers~\cite{doi:10.1098/rspa.1936.0215}. However, it is now understood that the observed subcritical transition was largely an endcap artifact. Removing endcap effects, Taylor--Couette experiments show that hydrodynamic Keplerian flows are nonlinearly stable (the effective viscosity remains at the molecular level) up to Reynolds number $\mathrm{Re}\sim 2\times 10^{6}$~\cite{Ji2006cr}. The stability has also been confirmed in numerical simulations (see, for example, Fig.~9 in Ref.~\cite{1998RvMP701B}).

\subsection{Stability arguments based on energy equations}
The stability of hydrodynamic Keplerian flows can be understood using the kinetic-energy equations (cf. Eqs.~58 and 59 in Ref.~\cite{1998RvMP701B}):
\begin{align}
    &\frac{\partial}{\partial t}\braket{ \tfrac{1}{2}\rho u_R^2 } = 2\Omega\braket{ \rho u_R u_\phi } - \epsilon_{\text{diss},R},\label{eq:kinetic_R}\\
    &\frac{\partial}{\partial t}\braket{ \tfrac{1}{2}\rho u_\phi^2 } = -2\Omega\braket{ \rho u_R u_\phi } + q\Omega\braket{ \rho u_R u_\phi } - \epsilon_{\text{diss},\phi}, \label{eq:kinetic_phi}
\end{align}
where the shear rate $q\equiv -d\ln\Omega/d\ln r$ is $3/2$ in Keplerian flow. Equation~\eqref{eq:kinetic_phi} shows that the Reynolds-stress term, proportional to $q\Omega$, can extract energy from the background shear to drive instability, while the Coriolis terms exchange radial and azimuthal momenta and provide a restoring effect. Owing to this restoring effect (ultimately tied to angular-momentum conservation), hydrodynamic Keplerian flows are nonlinearly energy-stable (see the Lyapunov argument in Appendix~\ref{sec:Lyapunov}).

With magnetic fields, the Maxwell-stress term in Eq.~\eqref{eq:magnetic_phi}, again proportional to $q\Omega$, extracts energy from the background shear without an opposing Coriolis counterpart, thereby opening a channel for instability (cf. Eqs.~84 and 86 in Ref.~\cite{1998RvMP701B}):
\begin{align}
    &\frac{\partial}{\partial t}\braket{ \tfrac{1}{2}\rho u_R^2 + \tfrac{1}{2}B_R^2 }
    = 2\Omega\braket{ \rho u_R u_\phi } - \epsilon_{\text{diss},R}, \label{eq:magnetic_R}\\
    &\frac{\partial}{\partial t}\braket{ \tfrac{1}{2}\rho u_\phi^2 + \tfrac{1}{2}B_\phi^2 }
    = -2\Omega\braket{ \rho u_R u_\phi } + q\Omega\braket{ \rho u_R u_\phi }
      + q\Omega\braket{ -B_R B_\phi } - \epsilon_{\text{diss},\phi}. \label{eq:magnetic_phi}
\end{align}
However, the shear term (the $\Omega$-effect) only creates $B_\phi$ from a pre-existing $B_R$; it does not regenerate $B_R$. For MHD turbulence to be sustained, a process that regenerates $B_R$ is required to maintain positive feedback between Eqs.~\eqref{eq:magnetic_R} and \eqref{eq:magnetic_phi}. This regeneration is provided by the magnetorotational instability (MRI), enabling a dynamo in magnetized Keplerian flows, as shown below.

\subsection{A linearized fluid-element view of the MRI}
To illustrate how the MRI operates, consider the linearized motion of a magnetized fluid element in a frame corotating at angular frequency $\Omega$, with displacements $(\xi_R,\xi_\phi)$ (cf. Eqs.~106 and 107 of Ref.~\cite{1998RvMP701B}):
\begin{align}
&\begin{cases}
\ddot\xi_R - 2\Omega \dot\xi_\phi = -\dfrac{\partial\Phi_{\rm eff}}{\partial \xi_R},\\[4pt]
\ddot\xi_\phi + 2\Omega \dot\xi_R = -\dfrac{\partial\Phi_{\rm eff}}{\partial \xi_\phi},
\end{cases}\\
&\Phi_{\rm eff}(\xi_R,\xi_\phi)
= \tfrac{1}{2}\!\left[\left(\bm{k}\!\cdot\!\bm{u}_{\rm A}\right)^{2}-2q\Omega^{2}\right]\!\xi_R^{2}
 + \tfrac{1}{2}\left(\bm{k}\!\cdot\!\bm{u}_{\rm A}\right)^{2}\xi_\phi^{2},
\label{eq:effective_potential}
\end{align}
where $(\bm{k}\!\cdot\!\bm{u}_{\rm A})^{2}$ measures the effective magnetic tension. The conservative part of the dynamics is captured by motion in the effective potential $\Phi_{\rm eff}$, while the Coriolis force provides a velocity-dependent coupling that is therefore non-potential. Three regimes follow (Table~\ref{tab:mri_stability}).

The magnetic field enables angular-momentum exchange via magnetic tension; as a result, fluid-element motion is governed by $\Phi_{\rm eff}$ rather than epicyclic restoring alone in the field-free limit, and becomes destabilized when the field is sufficiently weak.
\begin{table}[ht]
\centering
\caption{Stability versus magnetic-field strength in Keplerian shear. Colors show $\Phi_{\rm eff}$ (Eq.~\ref{eq:effective_potential}); the gray curve is the trajectory of a slightly displaced fluid element (red) about its guiding center (blue). Movie files are provided as attachments; code to reproduce them is available in Ref.~\cite{MRI_basics}.}
\label{tab:mri_stability}
\begin{tabular}{ccc}
\toprule
$B=0$ & weak $B$: $(\bm{k}\!\cdot\!\bm{u}_A)^2<2q\Omega^2$ & strong $B$: $(\bm{k}\!\cdot\!\bm{u}_A)^2>2q\Omega^2$ \\
\midrule
stable (Rayleigh) & unstable (MRI) & stable (MRI) \\
\includegraphics[height=38.5mm]{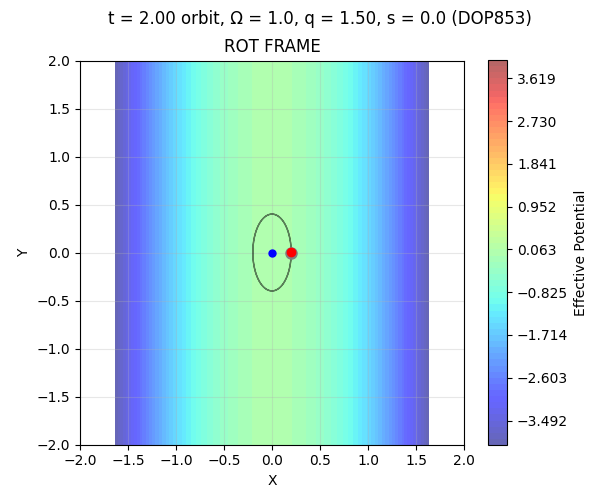} &
\includegraphics[height=38.5mm]{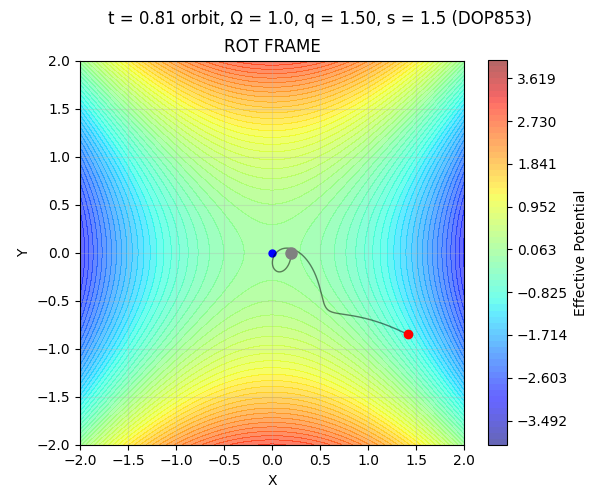} &
\includegraphics[height=38.5mm]{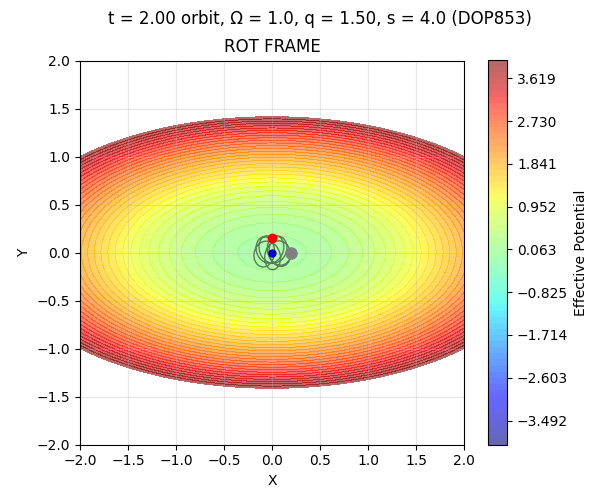} \\
\midrule
epicyclic motion &
$\Phi_{\rm eff}$ is a saddle &
$\Phi_{\rm eff}$ is confining \\
\bottomrule
\end{tabular}
\end{table}

\subsection{MRI turbulence and its saturation level}
Using MHD simulations, it has been confirmed that nonlinear MRI growth closes the dynamo loop and sustains MHD turbulence in Keplerian flows (e.g., Ref.~\cite{1998RvMP701B}). By contrast, MRI turbulence has not yet been achieved in laboratory experiments; for example, in Ref.~\cite{Wang2022}, other global MHD instabilities arising from conducting endcaps occurred before the MRI-unstable regime could be reached.

At present, numerical simulations are therefore the primary means to study the saturation level (the $\alpha$ value) of MRI turbulence. Ref.~\cite{Hawley_2011} showed that stratified shearing-box simulations with zero net flux exhibit robust saturation at $\alpha_{\rm MRI}\sim 0.02$, consistent across codes, resolutions, and box sizes. As pointed out by Refs.~\cite{10.1111/j.1365-2966.2007.11556.x,Kotko_2012}, this is much smaller than the hot-branch value inferred from observations, $\alpha_{\rm hot}\sim 0.1$. This discrepancy is notable because temperatures on the hot branch are high enough for MRI to operate.

\begin{figure}[ht]
    \centering
    \includegraphics[width=0.75\linewidth]{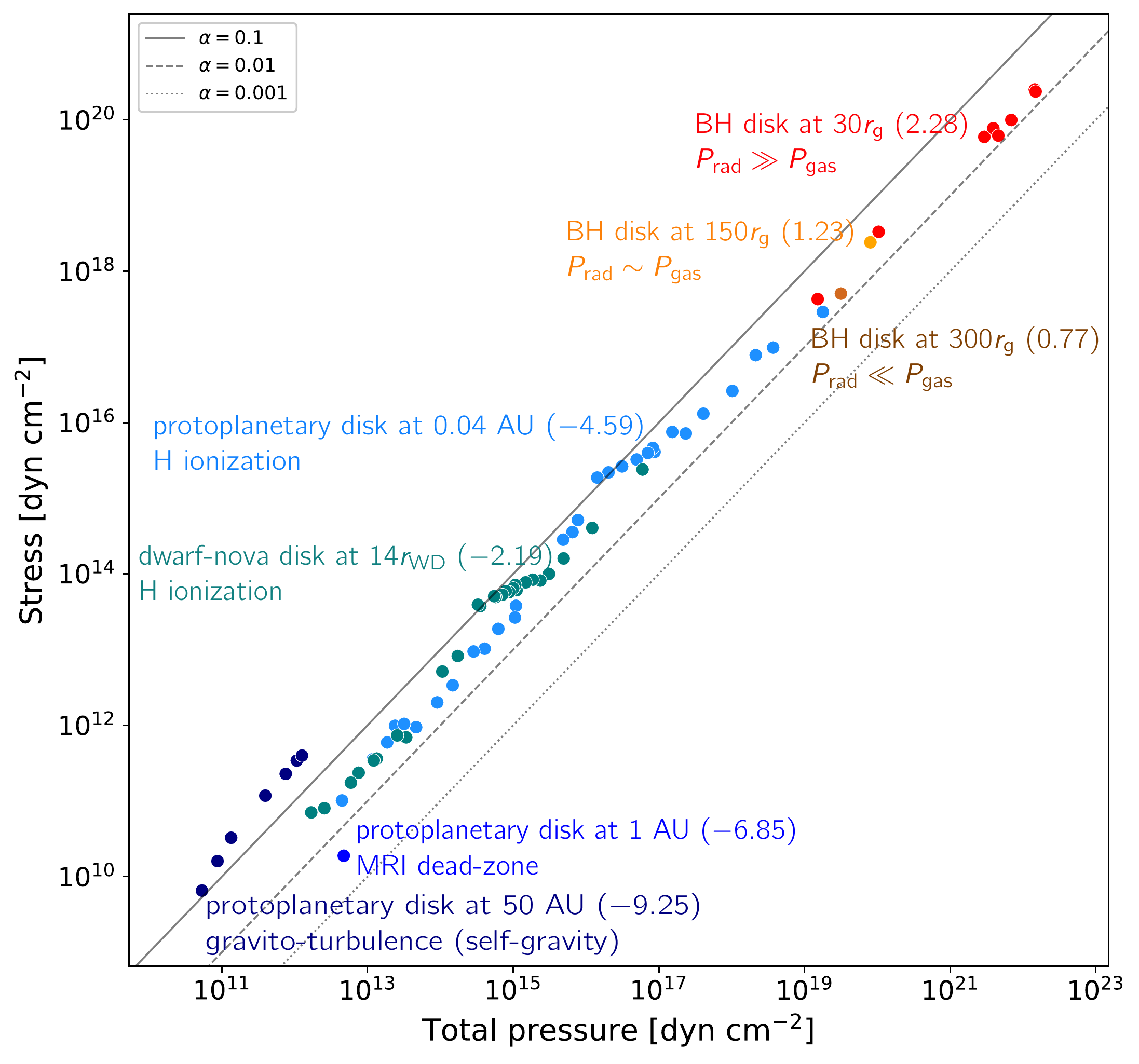}
    \caption{\rev{Shear stress versus total pressure ($P_{\rm gas} + P_{\rm rad}$)} for different disk classes; data from the references listed in the caption of Fig.~\ref{fig:TEcurve_all}. All data and scripts needed to reproduce this figure are available at Ref.~\cite{RMHD_TEs}.}
    \label{fig:universal_alpha}
\end{figure}

Although simulations in Ref.~\cite{Hawley_2011} employed simplified cooling, a similar near-universal value of $\alpha_{\rm MRI}\sim 0.02$ was also found in a series of radiation MHD simulations with realistic thermodynamics over nearly 12 orders of magnitude in pressure (Fig.~\ref{fig:universal_alpha}) ~\cite{Hirose_2006,Hirose_2009a,Hirose_2009b,Hirose_2011,Krolik_2007,Blaes_2007,Blaes_2011,Hirose_2014,10.1093/mnras/stv203,10.1093/mnras/stx824,10.1093/mnras/stz163}. There are a few exceptions: at outer radii of protoplanetary disks, gravito-turbulence (rather than MRI) dominates and $\alpha$ is much larger~\cite{10.1093/mnras/stx824,10.1093/mnras/stz163}; at intermediate radii, $\alpha$ is much smaller because weak ionization quenches MRI in the disk body~\cite{Hirose_2011}. Another exception is tied to hydrogen ionization in dwarf-nova disks~\cite{Hirose_2014} and in the inner protoplanetary disk~\cite{10.1093/mnras/stv203}, as discussed in the next section.

\section{Thermal equilibrium curves in MRI-driven dwarf-nova disks}\label{sec:s-curve}
\noindent A brief review of Ref.~\cite{Hirose_2014} is provided here, together with previously unpublished figures.

\subsection{Numerical Methods}
The ideal-MHD equations with radiative transfer were solved using the flux-limited diffusion (FLD) approximation and hydrogen-ionization thermodynamics, \rev{with precomputed tables for the EOS and the 
Rosseland- and Planck-mean opacities~\cite{optab14}}. A local shearing box represents a thin annulus at fixed radius; the angular velocity $\Omega$ sets the vertical gravity and inertia. The boundaries are shearing-periodic in $x$ (radial), periodic in $y$ (azimuthal), and outflow (free) in $z$. The mass in the box (i.e., $\Sigma$) is effectively fixed by the imposed vertical gravity. Because the $x$ boundaries are shearing-periodic, the domain is symmetric in $x$. Thus, the domain has no inner/outer edge across which mass can leave; even with finite stress, the net mass flux is zero. Consequently, dynamics and thermal evolution are captured, whereas viscous diffusion on long timescales is not.

For each chosen $\Omega$: (1) $\Sigma$ is selected and gas, radiation, and magnetic fields are initialized; (2) a quasi-steady state is evolved in which mechanical energy injection through the shearing boundaries balances the emergent radiative flux $F$ through the top and bottom, mediated by turbulent dissipation; (3) $T_{\rm eff}$ is inferred from $F = \sigma_{\rm B}T_{\rm eff}^4$; (4) steps (1)--(3) are repeated for different $\Sigma$ to obtain $T_{\rm eff}(\Sigma)$.

\begin{figure}[ht]
    \centering
    \includegraphics[width=0.7\linewidth]{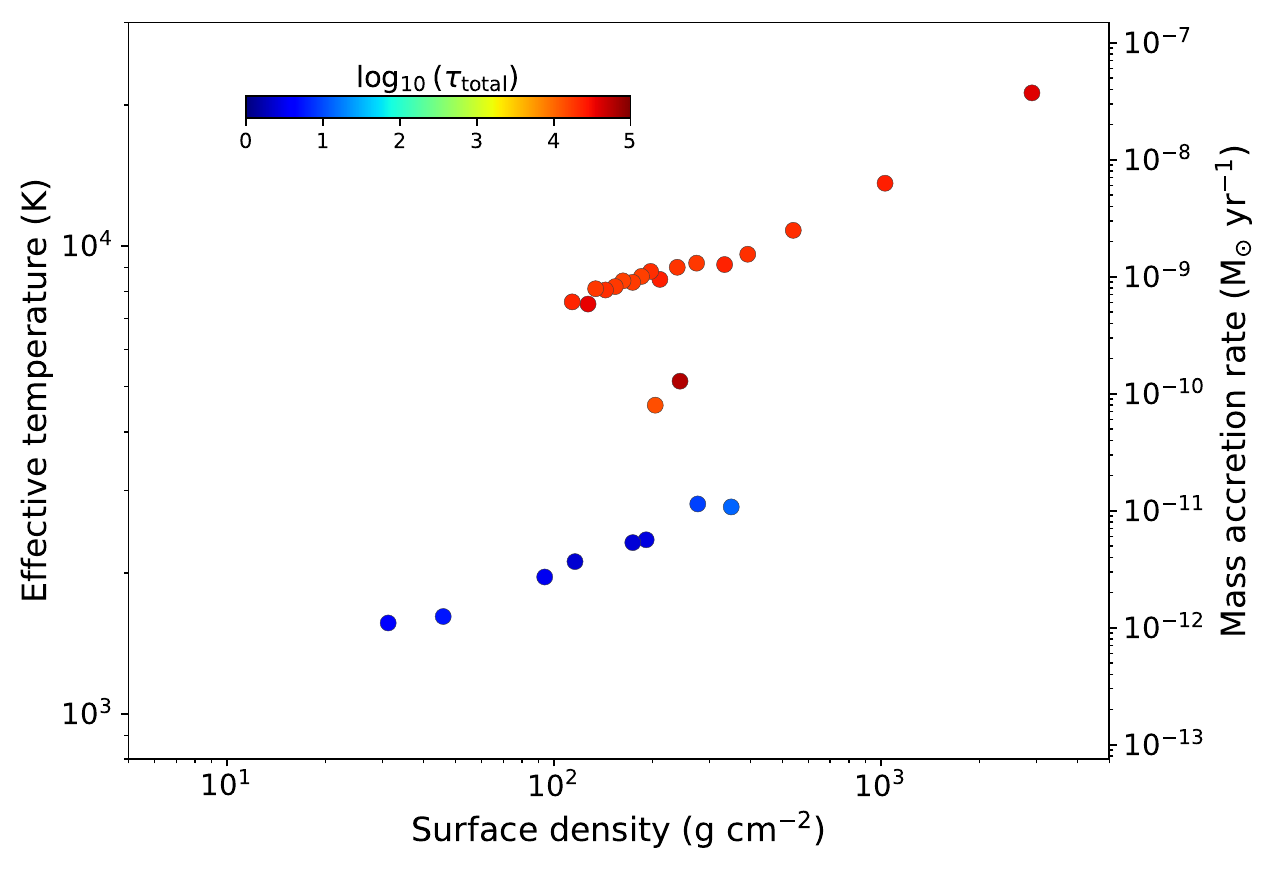}
    \caption{Thermal-balance equilibria at $\Omega = 6.4\times10^{-3}\ \mathrm{s}^{-1}$. The color indicates the total optical depth based on the Rosseland-mean opacity.}
    \label{fig:SC_tau}
\end{figure}

\subsection{First-principles S-curve near hydrogen-ionization temperatures}
Figure~\ref{fig:SC_tau} shows quasi-steady equilibria that populate an S-curve in the $T_{\rm eff}$--$\Sigma$ plane at $\Omega = 6.4\times10^{-3}\ \mathrm{s}^{-1}$. Color-coding by the total optical depth highlights an upper hot (optically thick) branch, a lower cool (optically thin) branch, and a sparse set of intermediate states. The sharp contrast in total optical depth between the hot and cool branches reflects hydrogen ionization, as discussed in Sec.~\ref{sec:section1}. 

\begin{figure}[ht]
    \centering
    \includegraphics[width=0.55\linewidth]{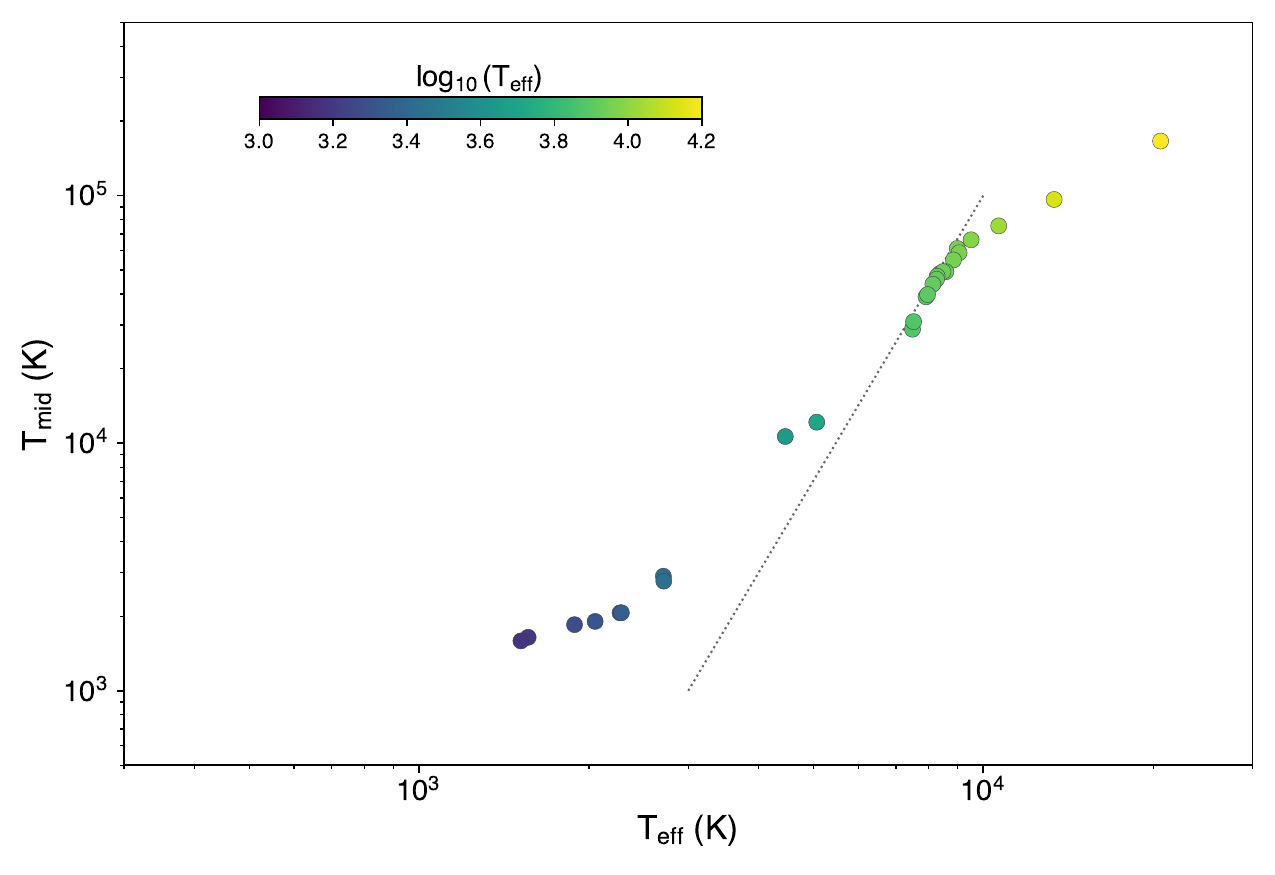}
    \caption{$T_{\rm mid}$ versus $T_{\rm eff}$. The dotted curve indicates $T_{\rm mid} \propto T_{\rm eff}^4$. The color encodes $T_{\rm eff}$.}
    \label{fig:Teff-Tmid}
\end{figure}

Because time integration converges only to stable equilibria, gaps separating the three branches imply existence of intervening unstable segments. Consistent with this, Fig.~\ref{fig:Teff-Tmid} (showing $T_{\rm eff}$ versus $T_{\rm mid}$) also indicates unstable intervals between branches. On the stable branches, the relation grows more slowly than the $T_{\rm eff}^{4}$ scaling, in line with vertical-structure expectations (cf. Eq.~\ref{eq:S-curve}).

\subsection{Vertical energy transport and flow morphology}
\begin{figure}[ht]
    \centering
    \includegraphics[width=0.7\linewidth]{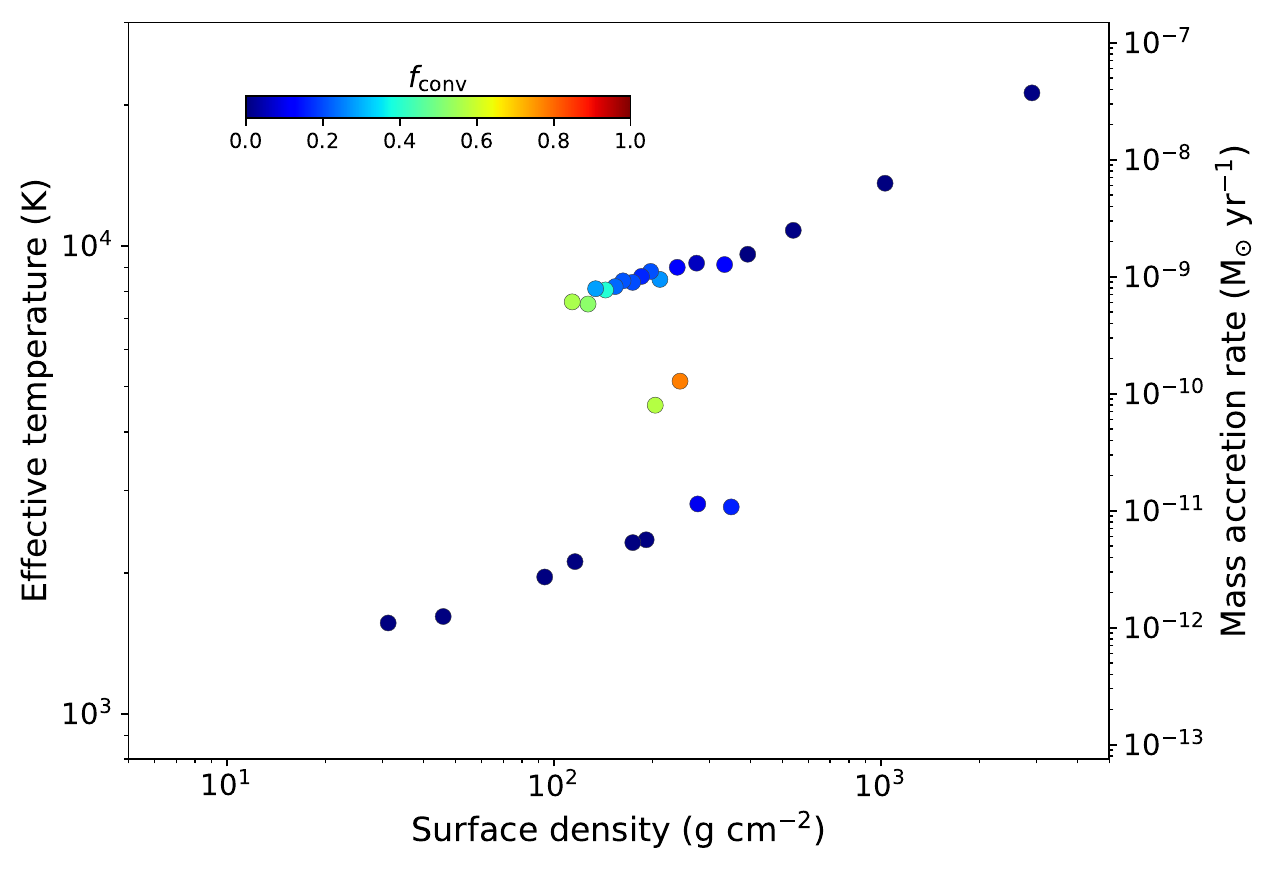}
    \caption{Same as Fig.~\ref{fig:SC_tau}, but the color indicates the advective fraction in the vertical energy transport; $f_{\rm conv} \equiv {\int\!\langle F_{\rm adv}\rangle\, dz}/({\int\!\braket{F_{\rm adv}}\,dz + \int\!\braket{F_{\rm rad}}\, dz})$, where $\braket{F_{\rm adv}}$ and $\braket{F_{\rm rad}}$ are time- and horizontally-averaged vertical advective and radiative energy fluxes, respectively.}
    \label{fig:SC_fadv}
\end{figure}

Figure~\ref{fig:SC_fadv} shows the vertical energy transport in the steady states; radiative cooling dominates vertical energy transport almost everywhere, except near the low-$\Sigma$ end of the upper branch and on the middle branch, where convective cooling carries a substantial fraction of the heat. 

\begin{figure}
    \centering
    \includegraphics[width=0.34\linewidth]{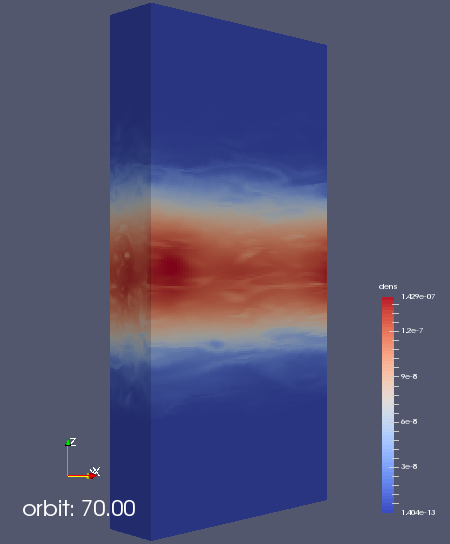}
    \includegraphics[width=0.34\linewidth]{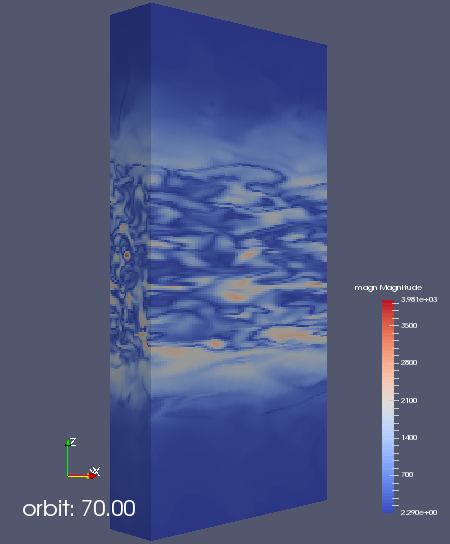}
    \caption{Snapshots of density (left) and magnetic-field strength (right) in a radiative case on the upper branch \rev{of the protoplanetary disk case ($\Omega = 2.6\times10^{-5}\ \mathrm{s}^{-1}$)}. Movie files are provided as attachments.}
    \label{fig:snapshot_rad}
\end{figure}

\begin{figure}
    \centering
    \includegraphics[width=0.34\linewidth]{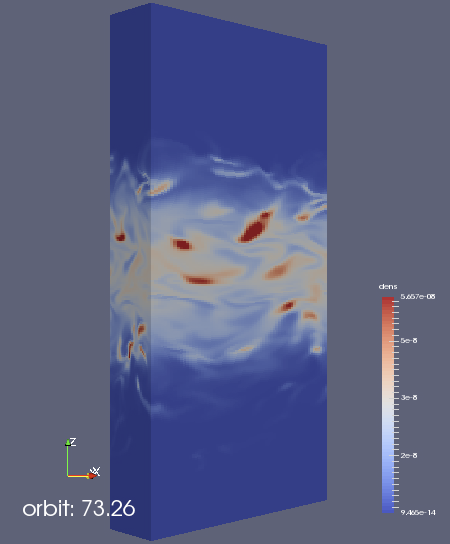}
    \includegraphics[width=0.34\linewidth]{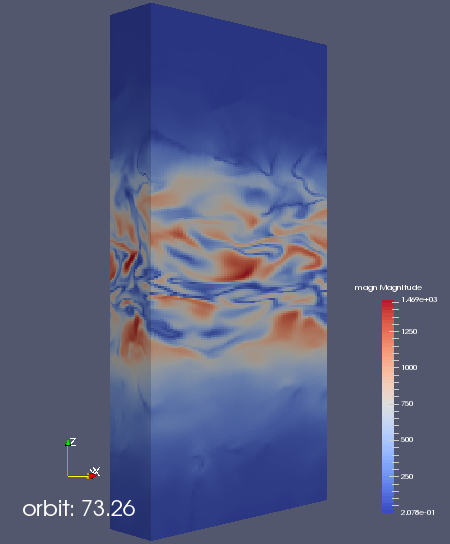}
    \caption{Same as Fig.~\ref{fig:snapshot_rad}, but for a convective case on the upper branch. Movie files are provided as attachments.}
    \label{fig:snapshot_conv}
\end{figure}

In radiative cases, motions are mostly horizontal as in standard MRI turbulence (Fig.~\ref{fig:snapshot_rad}). In convective cases, strong vertical motions appear: dense blobs fall toward the midplane and lighter blobs rise, exchanging places---a pattern characteristic of thermal convection (Fig.~\ref{fig:snapshot_conv}). The magnetic field follows the flow: in convective cases, vertical motions stretch the field and tilt it vertically. (For illustration, these snapshots use a case representative of \rev{a protoplanetary disk at 0.04 AU}~\cite{10.1093/mnras/stv203}; the underlying MRI-enhanced convection mechanism is the same as in dwarf-nova disks.)

\subsection{Convection-enhanced MRI and alpha}
\begin{figure}
    \centering
    \includegraphics[width=0.7\linewidth]{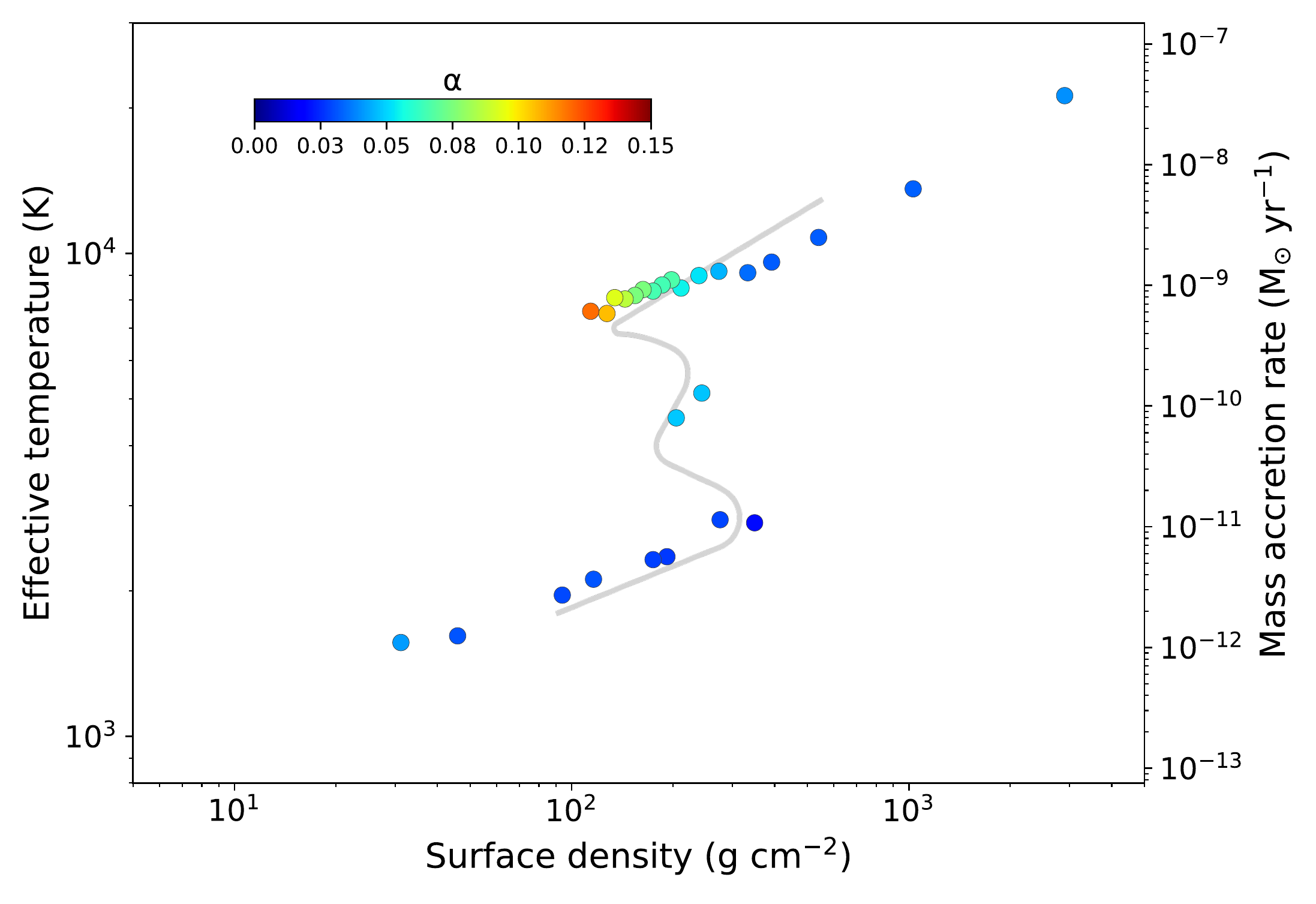}
    \caption{Same as Fig.~\ref{fig:SC_tau}, but the color indicates $\alpha$. The gray curve is a DIM model taken from \rev{Fig.~2 in Ref.~\cite{1985PASJ371M}, where $\Omega = 4.1\times10^{-3}\ \mathrm{s}^{-1}$, about 0.63 of that in the simulations.}}
    \label{fig:SC_alpha}
\end{figure}

The vertical stretching of the field in convective cases raises $\alpha$. Figure~\ref{fig:SC_alpha} shows that most steady states show the standard MRI level $\alpha \sim 0.02$, but near the low-$\Sigma$ end of the upper branch and on the middle branch, $\alpha$ is much higher; especially at the low-$\Sigma$ end it exceeds $0.1$. The enhancement has two parts: (i) vertical convection generates and strengthens vertical field, boosting the MRI dynamo and increasing the Maxwell stress; (ii) convection enhances cooling and lowers the pressure. Together, these increase $\alpha \equiv \text{(stress)}/\text{(pressure)}$.

Figure~\ref{fig:SC_alpha} also shows that the first-principles S-curve reproduces the DIM scales for outburst amplitude and duration that assume \rev{$\alpha_{\rm hot}=0.1$ and $\alpha_{\rm cool}=0.032$~\cite{1985PASJ371M}}. The same contrast arises in the simulations: convection-enhanced MRI accounts for the hot branch, whereas the cool branch reflects the standard MRI level.

\subsection{Summary of alpha on hot/cool branches}
Table~\ref{tab:alphas} summarizes hot- and cool-branch $\alpha$ values inferred from observations and simulations to date. The hot-branch $\alpha$ can be explained by convection-enhanced MRI found in the simulations. Although the cool-branch $\alpha$ also seems to be explained by standard MRI, the ideal-MHD assumption in Ref.~\cite{Hirose_2014} is not self-consistent, because low ionization can quench the MRI. Studies including Ohmic resistivity find that the MRI is indeed suppressed on the cool branch, even with a net vertical field, unless that field is sufficiently strong~\cite{Scepi_2018a, Scepi_2018b}. Therefore, the cool-branch value of $\alpha$ remains unresolved and is a key target for future work.

\begin{table}[ht]
\centering
\caption{Phenomenological $\alpha$ on the hot/cool branches vs.\ those obtained in shearing-box simulations.}\label{tab:alphas}
\begin{tabular}{ccccc}
\toprule
 &\makecell{phenom. $\alpha$} & \makecell{Hirose+ 14~\cite{Hirose_2014}\\[0pt]ideal MHD} & \makecell{Scepi+ 18a~\cite{Scepi_2018a}\\Ohmic resistivity} & \makecell{Scepi+ 18b~\cite{Scepi_2018b}\\Ohmic resistivity + $B_z$} \\[2mm]
\midrule
\midrule
\makecell{hot branch\\(ideal MHD\\[-4pt] regime)} & \makecell{0.1} & \makecell{0.1\\[-0pt]convection-\\[-4pt]enhanced MRI} & \makecell{0.1\\[-0pt]convection-\\[-4pt]enhanced MRI} & \makecell{0.1\\[-0pt]convection-\\[-4pt]enhanced MRI} \\
\hline
\makecell{cool branch\\(non-ideal \\[-4pt]MHD regime)} & 0.02 & \makecell{0.02\\standard MRI} & \makecell{0\\ quenched MRI} & \makecell{0\\ quenched MRI\\[-4pt](unless strong $B_z$)} \\
\bottomrule
\end{tabular}
\end{table}

\appendix

\section{Notes on stability of radiation-dominated accretion flows}\label{sec:notes_prad}
\noindent The discussion presented here is based on Refs.~\cite{Hirose_2009a,Hirose_2009b,Blaes_2011} and includes previously unpublished figures.

\subsection{Are radiation-dominated flows stable?}
The question divides into two parts: (1) Do such flows admit an equilibrium? (2) If so, is that equilibrium stable? These are not straightforward to answer for the following reasons.

{\color{black}First, the existence of an equilibrium is not self-evident. In radiation-dominated flows where vertical energy transport is by pure radiative diffusion with Thomson opacity $\kappa_{\rm T}$, hydrostatic balance implies $F(z) = \big(c\Omega^{2}/\kappa_{\rm T}\big)\,z$, so the volumetric cooling rate $q^{-}\!\equiv dF/dz$ is fixed at $c\Omega^{2}/\kappa_{\rm T}$~\cite{10.1093/mnras/175.3.613}; however, MRI turbulence---the most plausible heating source---has no known mechanism to self-tune its volumetric heating rate to that diffusion-set value.}

Numerically, questions (1) and (2) are entangled because runs start out of equilibrium; searching for a balance and testing its stability occur simultaneously. Moreover, a local equilibrium---if it exists---will be reachable only from initial states inside its basin of attraction.

\subsection{Quasi-steady states obtained in radiation MHD simulations}
\begin{figure}[ht]
    \centering
    \includegraphics[width=0.48\linewidth]{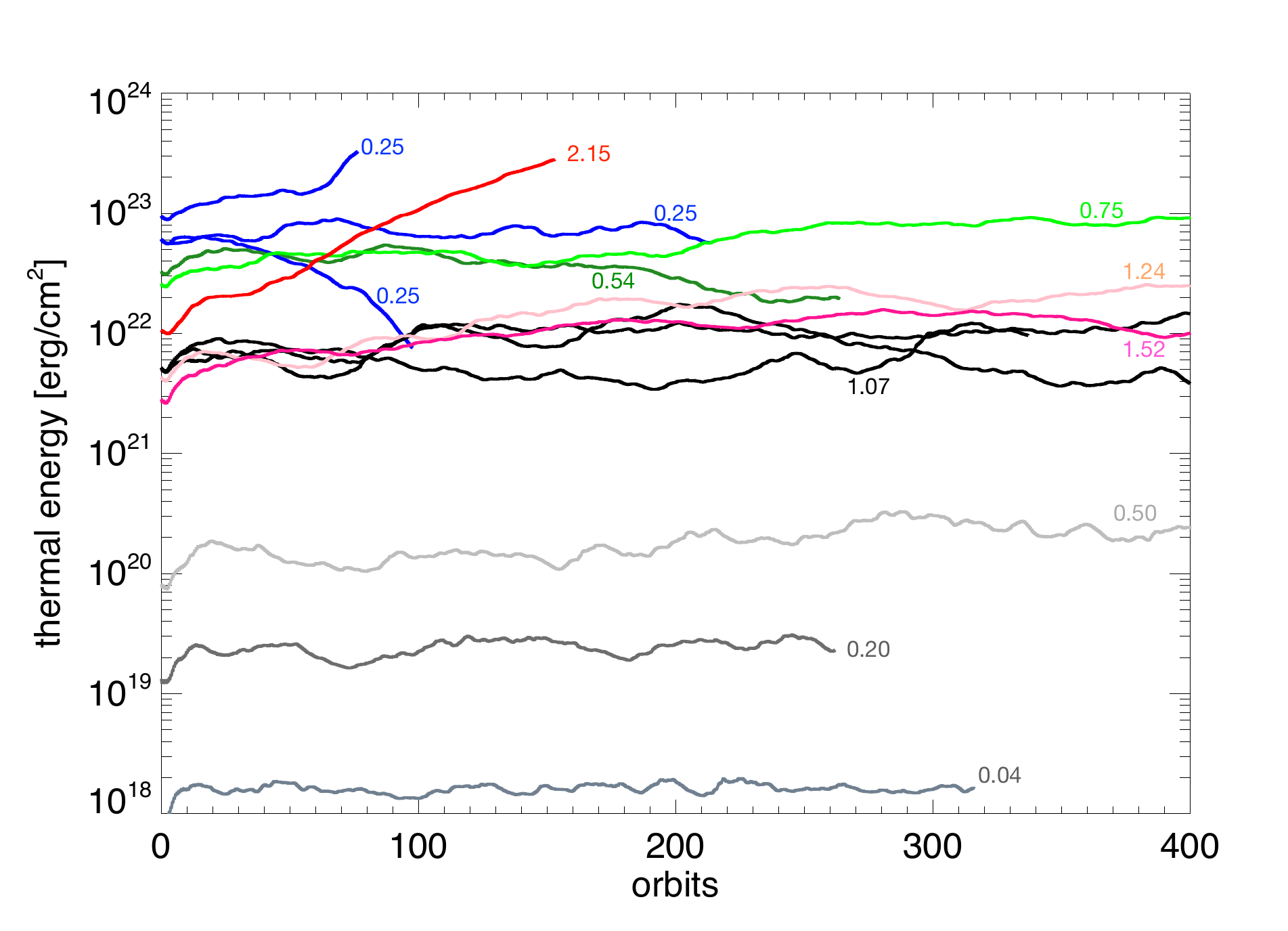}
    \includegraphics[width=0.48\linewidth]{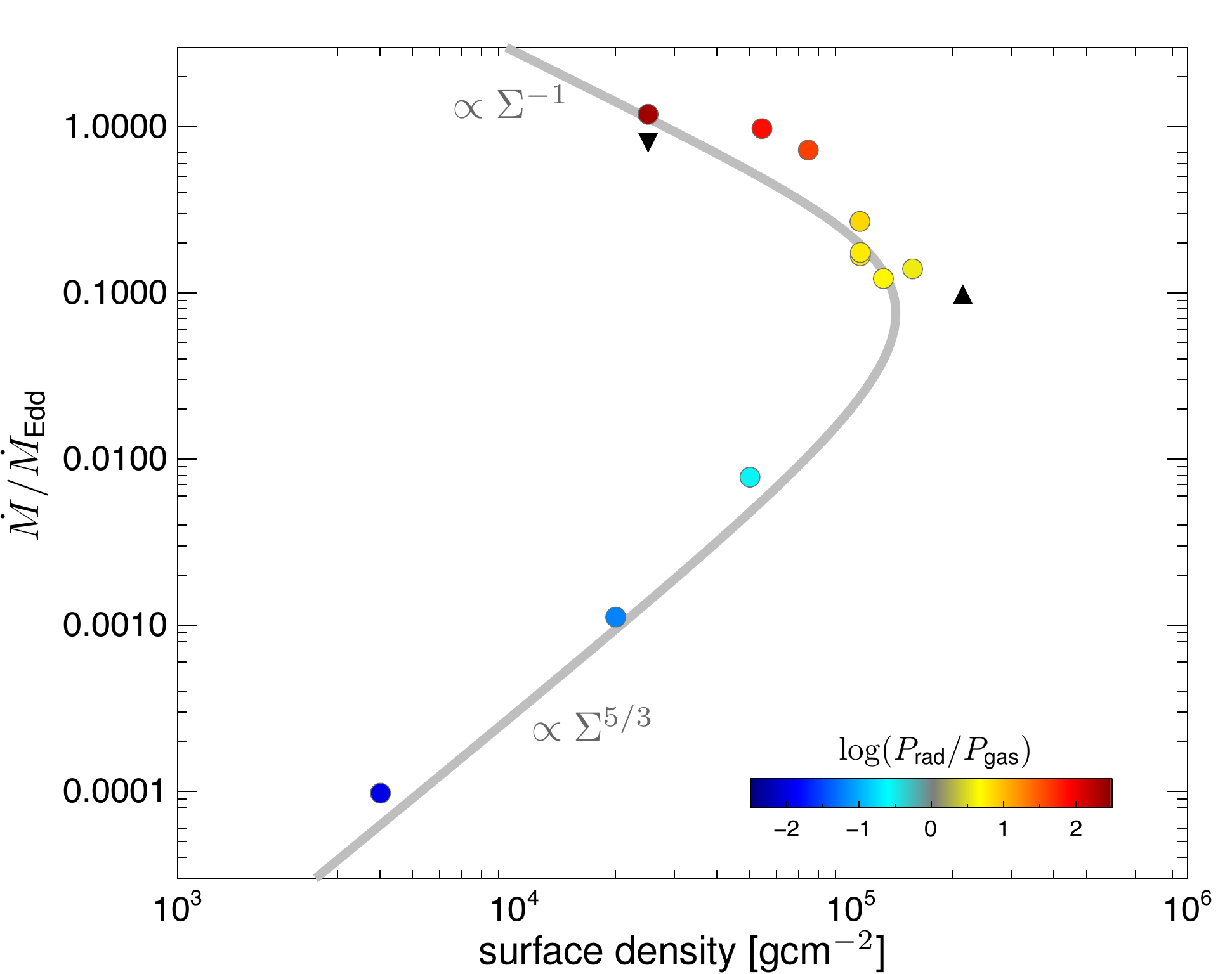}
    \caption{Left: Thermal histories at $\Omega = 190\ \mathrm{s}^{-1}$. Values beside the curves are surface densities in units of $10^5\ \mathrm{g\,cm^{-2}}$. Right: Quasi-steady states in the $(\Sigma,\dot M/\dot M_\mathrm{Edd})$ plane; the horizontal axis is $\Sigma$. The vertical axis assumes a radiative efficiency of $0.1$. Color shows $P_{\rm rad}/P_{\rm gas}$. Triangles mark runaway cases. The gray curve is a representative equilibrium locus for $\alpha=0.025$.}
    \label{fig:SC_prad}
\end{figure}

The $\alpha$-model with $Q^+\propto E$ and $Q^-\propto E^{1/2}$ for radiation-dominated disks predicts thermal instability~\cite{10.1093/mnras/175.3.613}. Yet Ref.~\cite{Hirose_2009a,Hirose_2009b,Blaes_2011} found radiation-dominated quasi-steady states lasting $\gtrsim 200$ orbits in the shearing-box simulations at $\Omega=190\ \mathrm{s}^{-1}$ (Fig.~\ref{fig:SC_prad}). In these states, vertical radiation advection coexists with diffusion so that (diffusion $+$ advection) $\approx$ heating in the mean, enabling a balance that pure diffusion alone would not~\cite{Hirose_2009a}. This addresses question (1), the existence of a local equilibrium.

How, then, can such a balance persist despite the $\alpha$-model prediction? One reason for this is that a dissipation lag reduces the in-phase {component} of the heating response to energy perturbations~\cite{Hirose_2009a}. In addition, vertical advection could play a role in stabilization; that is, small increases in thermal energy may enhance buoyant transport, thereby effectively steepening the cooling relative to pure diffusion. This remains a qualitative hypothesis, however, and is not quantified here.

Runaway behavior is also observed. At $\Sigma = 2.15\times10^5\ \mathrm{g\,cm^{-2}}$ a heating runaway occurs, while at $\Sigma = 0.25\times10^5\ \mathrm{g\,cm^{-2}}$ both heating and cooling runaways appear. The former suggests a turning point in the thermal-equilibrium locus beyond which no equilibrium exists, as predicted by the $\alpha$-model (gray curve in Fig.~\ref{fig:SC_prad}, right). The latter may indicate that radiation-dominated flows become genuinely \rev{thermally} unstable as the accretion rate approaches the Eddington value.
\rev{It should be noted that in this regime the viscous and thermal timescales become comparable, and thus variations in $\Sigma$ must also be considered---an effect not captured in the shearing-box approximation.}

\subsection{Comparison between ZEUS and Athena runs}

\begin{figure}[ht]
    \centering
    \includegraphics[width=0.48\linewidth]{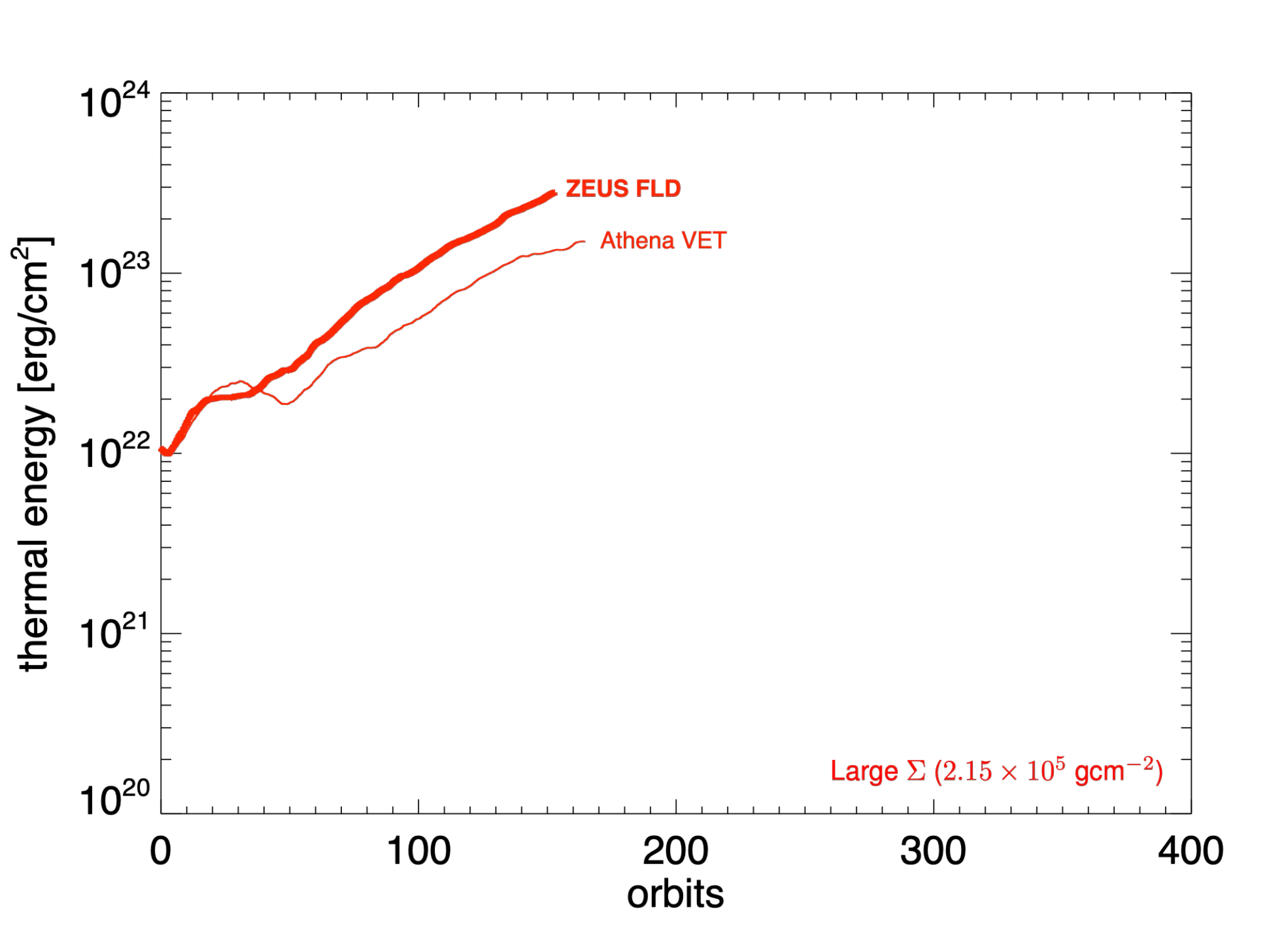}
    \includegraphics[width=0.48\linewidth]{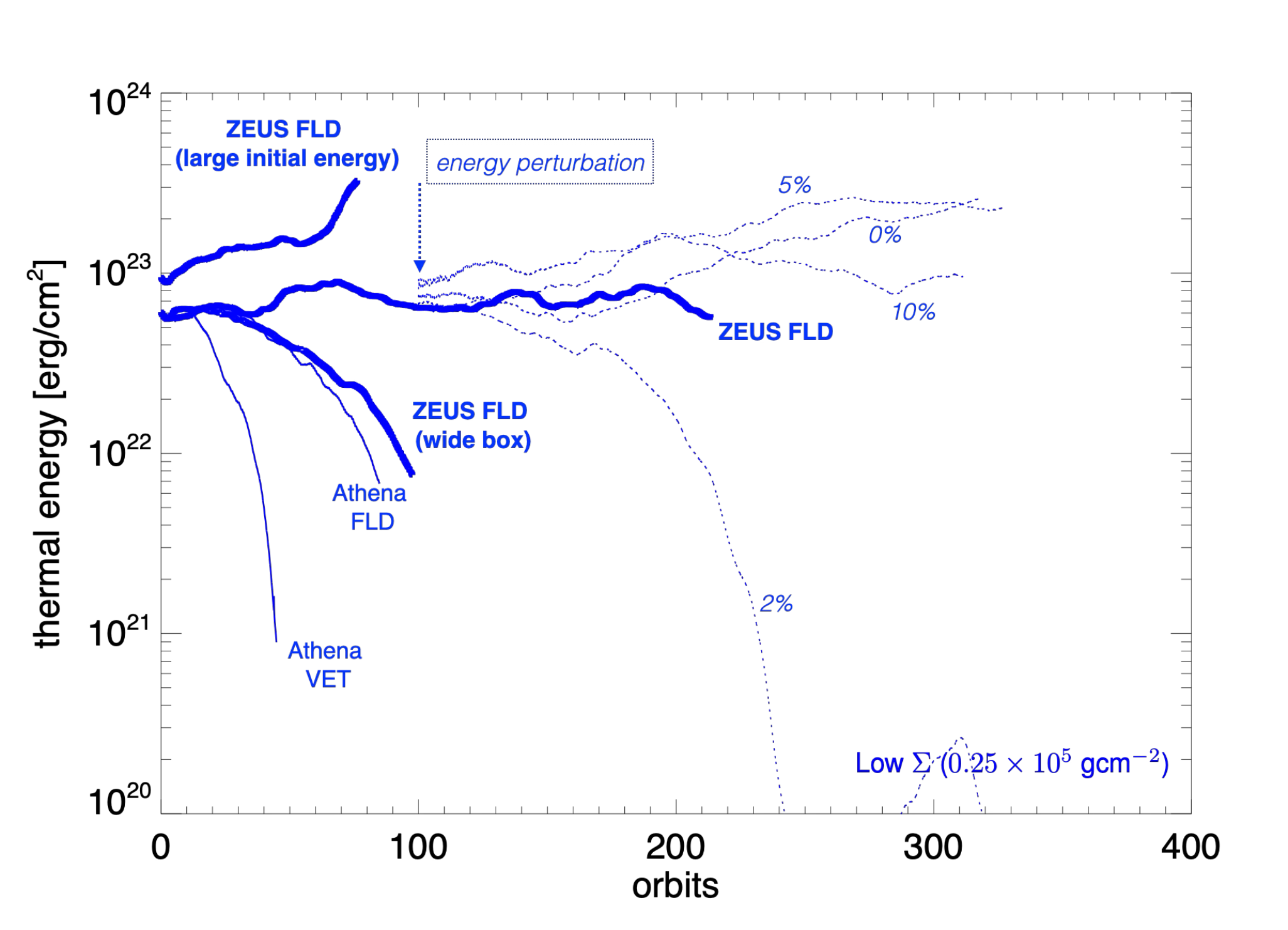}
    \includegraphics[width=0.48\linewidth]{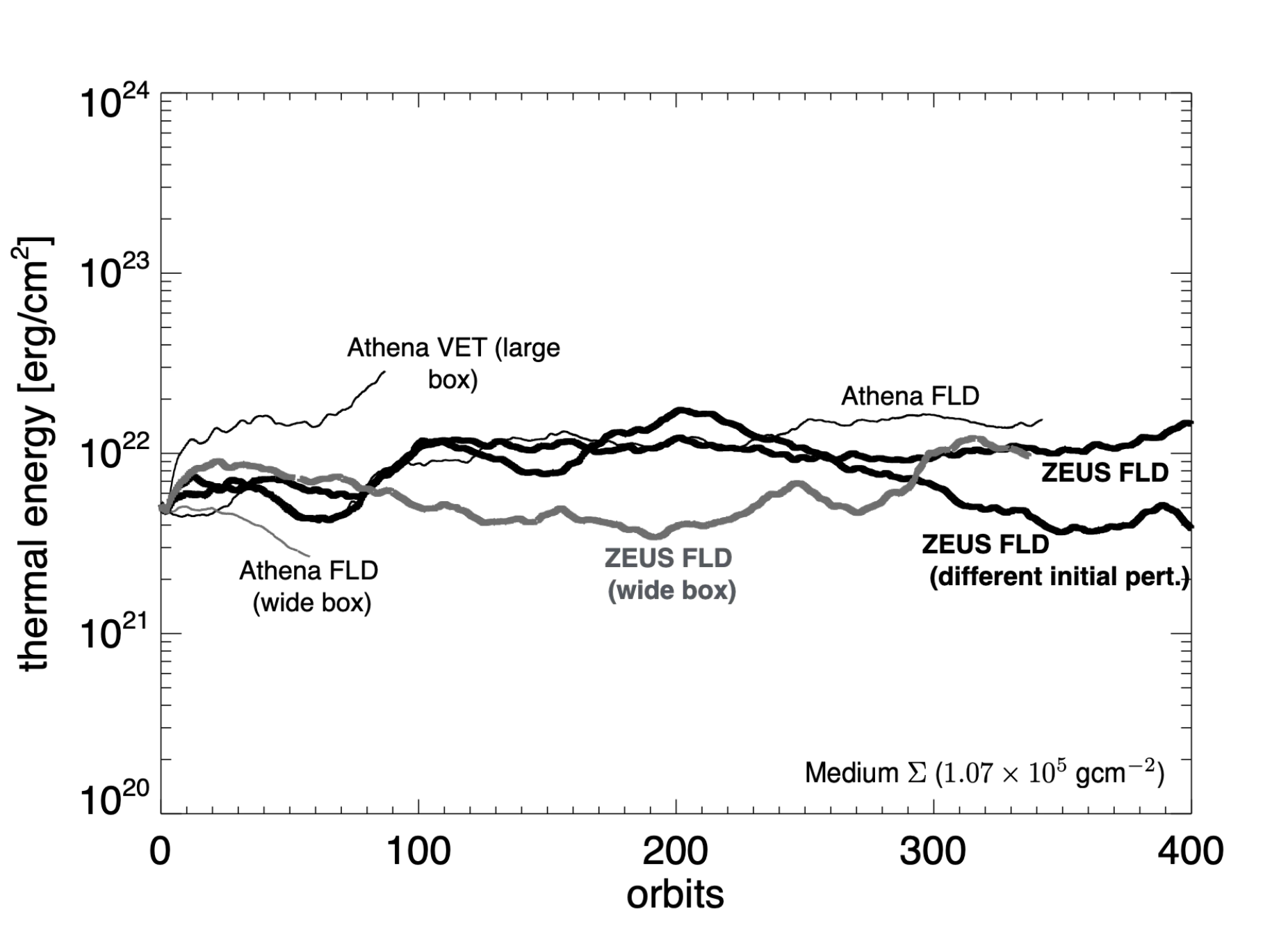}
    \includegraphics[width=0.48\linewidth]{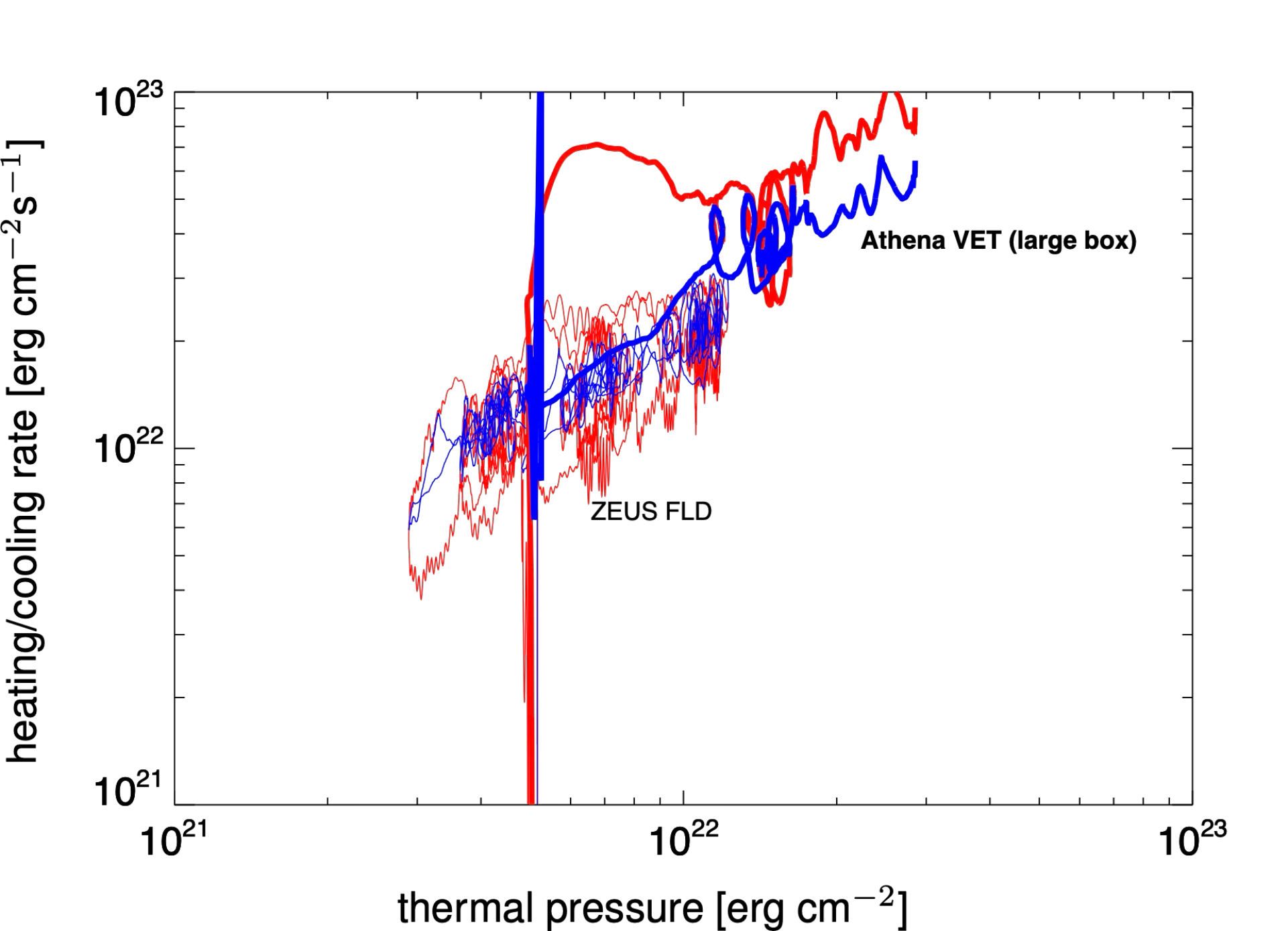}
    \caption{Thermal histories from ZEUS--FLD and Athena--VET/FLD runs at $\Sigma=2.15\times10^5\ \mathrm{g\,cm^{-2}}$ (top left), $0.25\times10^5\ \mathrm{g\,cm^{-2}}$ (top right), and $1.07\times10^5\ \mathrm{g\,cm^{-2}}$ (bottom left). Bottom right: trajectories of cooling (blue) and heating (red) versus thermal energy; thick lines: Athena--VET, thin lines: ZEUS--FLD. A movie file for the bottom-right panel (ZEUS--FLD) is provided as an attachment.}
    \label{fig:sigma_middle}
\end{figure}

Figure~\ref{fig:sigma_middle} compares ZEUS--FLD results~\cite{Hirose_2009a,Hirose_2009b,Blaes_2011} with Athena--VET (variable Eddington tensor)/FLD~\cite{Jiang_2013}, both of which simulated radiation-dominated accretion flows in the shearing-box framework. Because codes, closures, and grids differ, a one-to-one match is not expected, but some features agree: at $\Sigma=2.15\times10^5\ \mathrm{g\,cm^{-2}}$ (top left) both show runaway heating; at $\Sigma=0.25\times10^5\ \mathrm{g\,cm^{-2}}$ (top right), Athena--VET/FLD shows runaway cooling while ZEUS--FLD outcomes vary across setups, suggesting that high accretion-rate cases did not settle into balance, consistent with the discussion in the previous paragraph.

At $\Sigma = 1.07\times10^5\ \mathrm{g\,cm^{-2}}$, or $\dot M/\dot M_\mathrm{Edd}\sim 0.1$ (Fig.~\ref{fig:sigma_middle}, bottom left), ZEUS--FLD consistently attains quasi-steady states, whereas Athena--VET/FLD generally runs away (with one Athena--FLD exception). The contrast of $Q^+$/$Q^-$ variations (Fig.~\ref{fig:sigma_middle}, bottom right) shows an early, sharp $Q^+$ overshoot in Athena--VET that outpaces $Q^-$, versus milder, closely tracked $Q^+$ and $Q^-$ in ZEUS--FLD. A proximate cause of the divergent outcomes (runaway vs.\ quasi-steady) appears to be such differences in the early-time evolution (i.e., MRI growth from a simple seed field and transient gas--radiation adjustment), likely influenced by the numerical method including radiation closure, domain size, resolution, and boundary conditions. Convergence to a local equilibrium may also require initial states that lie within its basin of attraction. The available data do not show whether the basins differ between ZEUS--FLD and Athena--VET.

\section{Lyapunov proof of nonlinear stability in hydrodynamic Keplerian
flow}\label{sec:Lyapunov}
Consider the in-plane kinetic-energy equations (Eqs.~\ref{eq:kinetic_R} and \ref{eq:kinetic_phi}):
\begin{align}
\frac{d}{dt}\braket{ \tfrac12\rho u_R^2 }&=2\Omega\braket{\rho u_Ru_\phi}-\epsilon_{\mathrm{diss},R},\\
\frac{d}{dt}\braket{ \tfrac12\rho u_\phi^2 }&=(q-2)\Omega\braket{\rho u_Ru_\phi}-\epsilon_{\mathrm{diss},\phi}.
\end{align}
Let \(K\equiv\braket{\tfrac12\rho(u_R^2+u_\phi^2)}\) and \(\epsilon_{\mathrm{diss}}\equiv\epsilon_{\mathrm{diss},R}+\epsilon_{\mathrm{diss},\phi}\ge0\), so that
\begin{align}
\frac{dK}{dt}&=q\Omega\braket{\rho u_Ru_\phi}-\epsilon_{\mathrm{diss}}.\label{eq:dKdt}
\end{align}
Introduce \(L\equiv\braket{\tfrac12\rho(\alpha u_R^2+\beta u_\phi^2)}\) with constants \(\alpha,\beta>0\) chosen to satisfy
\begin{align}
2\alpha+(q-2)\beta&=-q.\label{eq:abcon}
\end{align}
For \(0<q<2\) (Rayleigh-stable), one convenient choice is \(\alpha=2-\tfrac{q}{2}>0\), \(\beta=\tfrac{4}{2-q}>0\). Taking \(dL/dt\) from the balances and using \eqref{eq:abcon} yields
\begin{align}
q\Omega\braket{\rho u_Ru_\phi}&=-\frac{dL}{dt}-\big(\alpha\,\epsilon_{\mathrm{diss},R}+\beta\,\epsilon_{\mathrm{diss},\phi}\big).\label{eq:injid}
\end{align}
Adding \eqref{eq:dKdt} and \eqref{eq:injid} gives the Lyapunov functional \(H\equiv K+L=\braket{\tfrac12\rho[(1+\alpha)u_R^2+(1+\beta)u_\phi^2]}\), whose time derivative is
\begin{align}
\frac{dH}{dt}&=-\left[(\alpha+1)\epsilon_{\mathrm{diss},R}+(\beta+1)\epsilon_{\mathrm{diss},\phi}\right]\le0.
\end{align}
Because \(0<q<2\) implies \(\alpha,\beta>0\), \(H\) is positive definite and nonincreasing for all finite-amplitude perturbations (all Reynolds numbers). Under standard bounded/periodic domains (Poincaré-type inequality), this implies \(K(t)\to0\) as \(t\to\infty\). Therefore the unmagnetized Keplerian flow (\(0<q<2\)) is nonlinearly energy-stable.

\acknowledgments
I thank the DIM 50th anniversary workshop organizers for the invitation and support.
I am grateful to my collaborators---J.H. Krolik, O. Blaes, N.J. Turner, J.-M. Shi, J.M. Stone, M.S.B. Coleman, T. Sano, and N. Shabaltas---for our earlier joint work (Refs.~\cite{Hirose_2009a,Krolik_2007,Blaes_2007,Hirose_2006,Hirose_2009b,Hirose_2011,Blaes_2011,Hirose_2014,10.1093/mnras/stv203,10.1093/mnras/stx824,10.1093/mnras/stz163}).
I also thank Y.-F. Jiang for sharing the data used in Appendix~\ref{sec:notes_prad}, \rev{and the anonymous referee for valuable comments that improved the manuscript.}

\bibliographystyle{JHEP}
\bibliography{references}


\end{document}